\documentclass[10pt,letterpaper]{article}
\usepackage{opex3}
\usepackage{cite}

\begin{document}

\title{The first- and second-order temporal interference between thermal and laser light}

\author{Jianbin Liu,$^{1,*}$ Huaibin Zheng,$^{1,2}$ Hui Chen,$^1$ Yu Zhou,$^2$ Fu-li Li,$^2$ and Zhuo Xu$^1$}

\address{$^1$Electronic Materials Research Laboratory, Key
Laboratory of the Ministry of Education \& International Center for
Dielectric Research, Xi'an Jiaotong University, Xi'an 710049, China \\$^2$MOE Key Laboratory for Nonequilibrium Synthesis and Modulation
of Condensed Matter, Department of Applied Physics,
Xi'an Jiaotong University, Xi'an 710049, China}

\begin{abstract}
The first- and second-order temporal interference between two independent thermal and laser light beams is discussed by employing the superposition principle in Feynman's path integral theory. It is concluded that the first-order temporal interference pattern can not be observed by superposing thermal and laser light, while the second-order temporal interference pattern can be observed in the same condition. These predictions are experimentally verified by employing pseudothermal light to simulate thermal light. The conclusions and method in the paper can be generalized to any order interference of light or massive particles, which is helpful to understand the physics of interference.
\end{abstract}



\section{Introduction}

Interference of light is essential to understand optical coherence theory \cite{glauber-1,glauber-2,mandel-rmp} and superposition principle in quantum physics \cite{dirac,feynman}. Based on the conservation of energy, Dirac concluded that \textit{``Each photon then interferes only with itself. Interference between two different photons never occurs.''} This simple and famous statement is the key to understand the first-order interference of light in quantum physics, which triggers lots of discussions about its correctness \cite{mandel-rmp,paul-rmp}. Mandel \textit{et al.} experimentally verified that there is transient first-order interference pattern by superposing two independent lasers \cite{mandel-nature}. However, after further experiments \cite{mandel-laser} and careful thinking,  they finally concluded that their experiments does not contradict Dirac's statement. The observed interference pattern is due to the detection of a photon \textit{``forces the photon into a superposition state in which it is partly in each beam. It is the two components of the state of one photon which interference, rather than two separate photons} \cite{mandel-rmp}''. Paul pointed out that the second part of Dirac's statement is not always correct by taking the second- and higher-order interference of light into consideration \cite{paul-rmp}. Dirac's statement is for the first-order interference of light. When considering the second- and higher-order interference of light, Dirac's statement should be generalized to that a multi-photon state of independent photons only interferes with itself and interference between two different multi-photon states never occurs \cite{shih-book}. From this point of view, Dirac's statement has not been disproved when considering the second- and higher-order interference. However, there are still different opinions about Dirac's statement in the optical community.

There is an alternative interpretation for the interference of light to avoid the discussions about Dirac's statement, which is the superposition principle in Feynman's path integral theory \cite{feynman}. For instance, there are two different ways to trigger a photon detection event in the first-order interference of two independent lasers \cite{mandel-nature}, which are emitted by the superposed two lasers, respectively. If these two different ways are indistinguishable, the probability distribution for the $j$th detected photon is given by \cite{feynman}
\begin{equation}
P_j(\vec{r},t)=|A_{j1}(\vec{r},t)+A_{j2}(\vec{r},t)|^2,
\end{equation}
where $A_{j1}(\vec{r},t)$ and $A_{j2}(\vec{r},t)$ are the probability amplitudes of the detected photon at $(\vec{r},t)$ is emitted by laser one and two, respectively. It does not matter whether these two different amplitudes belong to one photon or different photons. What matters is the indistinguishability of these two different alternatives. If these different alternatives are indistinguishable, there is interference. Otherwise, there is no interference \cite{feynman}. This principle is valid not only for the first-order interference, but also for the second- and higher-order interference. With the superposition principle in Feynman's path integral theory, one can not only avoid the discussions about Dirac's statement, but also get a unified interpretation for the interference of light.

The first- and second-order interference between two independent light beams has been studied extensively by employing different kinds of light sources, such as lasers \cite{javan,mandel-nature,lipsett,mandel-laser,kaltenbaek,liu-laser,kim-2014}, thermal light sources \cite{martienssen,ou,zhai,ou-2010,nevet,chen,liu-OE}, and nonclassical light sources \cite{mandel-rmp,paul-rmp,mandel-rmp-1999,zeilinger-rmp,pan-rmp}. The interference between photons in different kinds of light seems more interesting and important to understand the coherence properties of light, such as the interference of photons emitted by laser and single-photon sources \cite{bennet}, laser and quantum down-converted light \cite{afek}, and laser and thermal light beams \cite{liu-EPL,liu-arXiv}. Based on our recent studies about the second-order interference between thermal and laser light \cite{liu-EPL,liu-arXiv}, we will study the first- and second-order temporal interference between thermal and laser light by employing the superposition principle in Feynman's path integral theory, hoping to understand coherence properties of thermal and laser light better. We also studied the second-order temporal interference between thermal and laser light beams when the polarizations of the superposed light beams are different.

The following parts are organized as follows. In Sect. \ref{theory}, we will theoretically study the first- and second-order temporal interference between thermal and laser light based on the superposition principle in Feynman's path integral theory. The experiments are presented in Sect. \ref{experiments} by employing pseudothermal light to simulate thermal light. The discussions and conclusions are in Sects. \ref{discussions} and \ref{conclusions}, respectively.

\section{Theory}\label{theory}

\begin{figure}[htb]
    \centering
    \includegraphics[width=80mm]{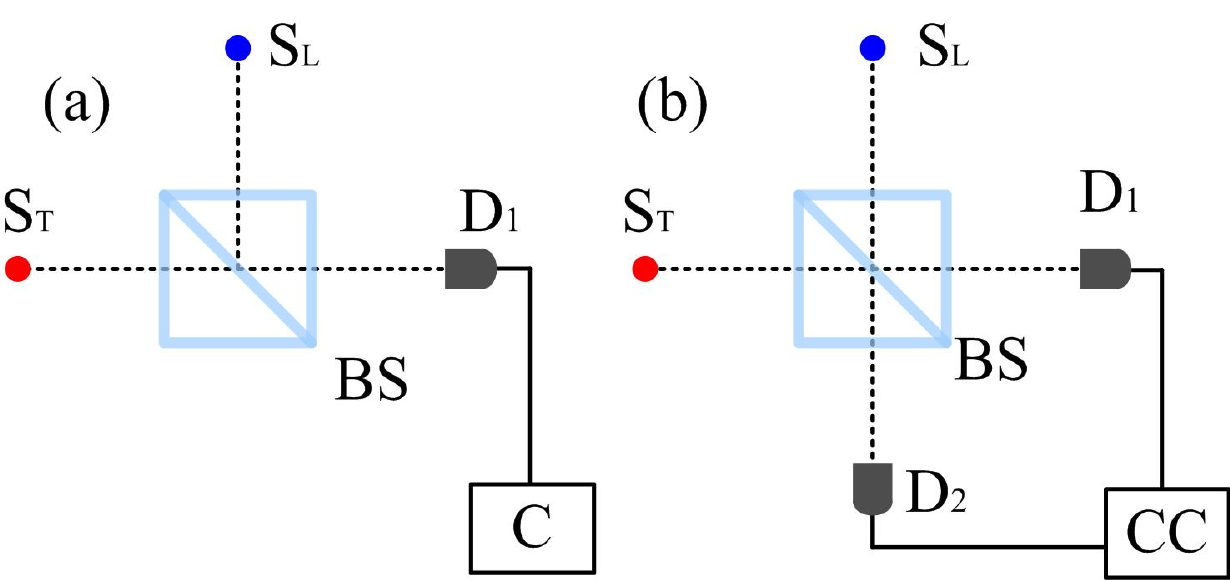}
    \caption{The first- and second-order interference between thermal and laser light beams. S$_T$ and S$_L$ are point thermal and laser light sources, respectively. D$_1$ and D$_2$ are two single photon detectors.  BS: 1:1 non-polarized beam splitter. C: single-photon count detection system. CC: two-photon coincidence count detection system.
    }\label{setup}
\end{figure}

We will employ the scheme in Fig. \ref{setup} for the calculations of the first- and second-order interference between thermal and laser light. Two independent thermal and laser light beams are incident to the two adjacent input ports of a 1:1 non-polarized beam splitter (BS), respectively. S$_T$ and S$_L$ are point thermal and laser light sources, respectively. D$_1$ and D$_2$ are two single photon detectors. The distance between the source and detection planes are all equal. C and CC are single-photon count and two-photon coincidence count detection systems, respectively. For simplicity, the polarizations and intensities of these two light beams are assumed to be the same. The mean frequencies of light emitted by S$_L$ and S$_T$ are $\omega_L$ and $\omega_T$, respectively.

In the first-order interference between thermal and laser light beams shown in Fig. \ref{setup}(a), there are two different alternatives to trigger a photon detection event at D$_1$. One is the detected photon is emitted by S$_L$. The other one is the detected photon is emitted by S$_T$. Although the frequencies of the photons emitted by S$_L$ and S$_T$  are different, these two different alternatives are indistinguishable if the time measurement uncertainty of the detection system is less than $1/|\omega_L-\omega_T|$ \cite{liu-arXiv} (and references therein). The probability distribution for the $j$th detected photon is given by \cite{feynman}
\begin{equation}\label{1st-1}
P_j^{(1)}(\vec{r},t)=|e^{i(\varphi_{Lj}+\pi/2)}K_{L1}(\vec{r},t)+e^{i\varphi_{Tj}}K_{T1}(\vec{r},t)|^2,
\end{equation}
where $\varphi_{Lj}$ and $\varphi_{Tj}$ are the initial phases of the $j$th detected photon emitted by S$_L$ and S$_T$, respectively. $K_{L1}(\vec{r},t)$ and $K_{T1}(\vec{r},t)$ are the Feynman's photon propagators from the S$_L$ and S$_T$ to D$_1$ at $(\vec{r},t)$, respectively. The extra phase $\pi/2$ is due to the photon reflected by the beam splitter will gain an extra phase comparing to the transmitted one \cite{loudon}. The magnitudes of these two amplitudes in Eq. (\ref{1st-1}) are equal is because the intensities of these two light beams are assumed to be identical. If the intensities are not equal, the magnitudes of these two amplitudes will be different.

The first-order interference pattern is proportional to the final probability distribution, which is equal to the sum of all the detected single-photon probability distributions
\begin{equation}\label{1st-2}
P^{(1)}(\vec{r},t)=\sum_jP^{(1)}_j(\vec{r},t)\equiv \langle |e^{i(\varphi_{Lj}+\pi/2)}K_{L1}(\vec{r},t)+e^{i\varphi_{Tj}}K_{T1}(\vec{r},t)|^2 \rangle,
\end{equation}
where $\langle...\rangle$ is ensemble average by taking all the detected single-photon distributions into consideration.
Photons in thermal light are emitted by spontaneous emissions and the initial phases of photons in thermal light are random. Photons in laser light are emitted by stimulated emissions and the initial phases are identical within the coherence time \cite{einstein}. Since the photons in thermal and laser light are independent, $\langle e^{i(\varphi_{Lj}-\varphi_{Tj})}\rangle$ equals 0. Equation (\ref{1st-2}) can be simplified as
\begin{equation}\label{1st-3}
P^{(1)}(\vec{r},t)=\langle |K_{L1}(\vec{r},t)|^2 \rangle +\langle |K_{T1}(\vec{r},t)|^2 \rangle,
\end{equation}
which is a constant. No first-order interference pattern can be observed.

In the second-order interference between thermal and laser light shown in Fig. \ref{setup}(b), there are four different cases to trigger a two-photon coincidence count. The first one is both photons are emitted by S$_T$. The second one is both photons are emitted by S$_L$. The third one is photon A is emitted by S$_T$ and photon B is emitted by S$_L$. The fourth one is photon A is emitted by S$_L$ and photon B is emitted by S$_T$. In the first case, there are two different alternatives to trigger a two-photon coincidence count, which are $A\rightarrow D_1, B\rightarrow D_2$ and $A\rightarrow D_2, B\rightarrow D_1$, respectively. $A\rightarrow D_1$ is short for photon A goes to detector 1 (D$_1$) and other symbols are defined similarly. In the second case, both photons are emitted by laser source, S$_L$. There should be two alternatives, too. However, there is only one alternative since these two alternatives are identical \cite{liu-arXiv}. In the third case, there are two alternatives, which are $A\rightarrow D_1, B\rightarrow D_2$ and $A\rightarrow D_2, B\rightarrow D_1$, respectively. The fourth case is similar as the third one. If all the seven different alternatives are indistinguishable, the $j$th detected two-photon probability distribution is \cite{feynman,liu-arXiv}
\begin{eqnarray}\label{G2-1}
&&P_j^{(2)}(\vec{r}_1,t_1;\vec{r}_2,t_2)\nonumber\\
&=&  |e^{i\varphi_{TjA}}K_{T1}e^{i(\varphi_{TjB}+\frac{\pi}{2})}K_{T2}+e^{i(\varphi_{TjA}+\frac{\pi}{2})}K_{T2}e^{i\varphi_{TjB}}K_{T1} \nonumber \\
&+& e^{i(\varphi_{Lj}+\frac{\pi}{2})}K_{L1}e^{i\varphi_{Lj}}K_{L2} \nonumber\\
&+& e^{i\varphi_{TjA}}K_{T1}e^{i\varphi_{Lj}}K_{L2}+e^{i(\varphi_{TjA}+\frac{\pi}{2})}K_{T2}e^{i(\varphi_{Lj}+\frac{\pi}{2})}K_{L1} \nonumber \\
&+& e^{i\varphi_{TjB}}K_{T1}e^{i\varphi_{Lj}}K_{L2}+e^{i(\varphi_{TjB}+\frac{\pi}{2})}K_{T2}e^{i(\varphi_{Lj}+\frac{\pi}{2})}K_{L1}|^2 .
\end{eqnarray}
Where $\varphi_{TjA}$ and $\varphi_{TjB}$ are the initial phases of photons A and B emitted by thermal source in the $j$th detected photon pair, respectively. $\varphi_{Lj}$ is the initial phase of photon emitted by laser in the $j$th detected photon pair. $K_{\alpha\beta}$ is short for $K_\alpha(\vec{r}_\beta,t_\beta)$, which means the Feynman's photon propagator from the light source S$_\alpha$ to the detector $\beta$ at $(\vec{r}_\beta,t_\beta)$ ($\alpha=L$ and T, $\beta=1$ and 2). The final two-photon probability distribution is the sum of all the detected two-photon probability distributions,
\begin{eqnarray}\label{G2-2}
&&P^{(2)}(\vec{r}_1,t_1;\vec{r}_2,t_2)=\sum_jP_j^{(2)}(\vec{r}_1,t_1;\vec{r}_2,t_2) \nonumber\\
&\equiv&\langle  |e^{i\varphi_{TjA}}K_{T1}e^{i(\varphi_{TjB}+\frac{\pi}{2})}K_{T2}+e^{i(\varphi_{TjA}+\frac{\pi}{2})}K_{T2}e^{i\varphi_{TjB}}K_{T1} \nonumber \\
&&+ e^{i(\varphi_{Lj}+\frac{\pi}{2})}K_{L1}e^{i\varphi_{Lj}}K_{L2} \nonumber\\
&&+ e^{i\varphi_{TjA}}K_{T1}e^{i\varphi_{Lj}}K_{L2}+e^{i(\varphi_{TjA}+\frac{\pi}{2})}K_{T2}e^{i(\varphi_{Lj}+\frac{\pi}{2})}K_{L1} \nonumber \\
&&+ e^{i\varphi_{TjB}}K_{T1}e^{i\varphi_{Lj}}K_{L2}+e^{i(\varphi_{TjB}+\frac{\pi}{2})}K_{T2}e^{i(\varphi_{Lj}+\frac{\pi}{2})}K_{L1}|^2 \rangle ,
\end{eqnarray}
where $\langle...\rangle$ is ensemble average by taking all the detected two-photon probability distributions into consideration. The four lines on the righthand side of Eq. (\ref{G2-2}) correspond to four different cases above, respectively. Since photons in thermal and laser light are independent, $\langle e^{i\varphi_{TjA}-\varphi_{TjB}} \rangle$, $\langle e^{i\varphi_{TjA}-\varphi_{Lj}} \rangle$, and $\langle e^{i\varphi_{TjB}-\varphi_{Lj}} \rangle$ all equal 0. Equation \ref{G2-2} can be simplified as
\begin{eqnarray}\label{G2-3}
&&P^{(2)}(\vec{r}_1,t_1;\vec{r}_2,t_2)\nonumber\\
&=& \langle |K_{T1}K_{T2}+K_{T2}K_{T1}|^2 \rangle \nonumber \\
&&+ \langle |K_{L1}K_{L2}|^2 \rangle \nonumber\\
&&+ \langle |K_{T1}K_{L2}-K_{T2}K_{L1}|^2 \nonumber \\
&&+ \langle |K_{T1}K_{L2}-K_{T2}K_{L1}|^2 \rangle .
\end{eqnarray}
The first line on the righthand side of Eq. (\ref{G2-3}) is two-photon bunching of thermal light \cite{HBT}. The second line corresponds to two-photon probability distribution of single-mode continuous-wave laser light, which is a constant \cite{glauber-1,glauber-2}. The third and fourth lines are two-photon beating terms when two photons are emitted by two light sources, respectively.

For a point light source, Feynman's photon propagator is \cite{peskin}
\begin{equation}\label{green}
K_{\alpha\beta}=\frac{\exp[-i(\vec{k}_{\alpha\beta}\cdot\vec{r}_{\alpha\beta}-\omega_{\alpha} t_{\beta})]}{r_{\alpha\beta}},
\end{equation}
which is the same as Green function in classical optics \cite{born}. $\vec{k}_{\alpha\beta}$ and $\vec{r}_{\alpha\beta}$ are the wave and position vectors of the photon emitted by S$_\alpha$ and detected at D$_\beta$, respectively. $r_{\alpha\beta}=|\mathbf{r}_{\alpha\beta}|$ is the distance between S$_\alpha$ and D$_\beta$. $\omega_{\alpha}$ and $t_{\beta}$ are the frequency and time for the photon that is emitted by S$_\alpha$ and detected at D$_\beta$, respectively ($\alpha=L$ and $T$, $\beta=1$ and 2). We will concentrate on the first- and second-order temporal interference between thermal and laser light. Generalizing the discussions to the spatial part is straight forward. Substituting Eq. (\ref{green}) into Eq. (\ref{G2-2}) and with similar calculations as the ones in Refs. \cite{liu-laser,liu-OE,liu-EPL,liu-pra,shih-book}, it is straight forward to have one-dimension temporal two-photon probability distribution as
\begin{eqnarray}\label{G2-4}
&&P^{(2)}(t_1-t_2)\nonumber\\
&\propto & 7+{sinc}^2\frac{\Delta\omega_T(t_1-t_2)}{2}\nonumber\\
&&-4\cos[\Delta\omega_{TL}(t_1-t_2)]{sinc}\frac{\Delta\omega_T(t_1-t_2)}{2}.
\end{eqnarray}
Where paraxial and quasi-monochromatic approximations have been employed to simplify the calculations. The positions of D$_1$ and D$_2$ are the same in order to concentrate on the temporal part. $sinc(x)$ equals $\sin x/x$. $\Delta\omega_T$ is the frequency bandwidth of thermal light. $\Delta\omega_{TL}$ is the difference between the mean frequencies of thermal and laser light. When the mean frequencies of thermal and laser light are different, the second-order temporal beating can be observed as shown by the last term of Eq. (\ref{G2-4}).

\section{Experiments}\label{experiments}

\begin{figure}[htb]
    \centering
    \includegraphics[width=60mm]{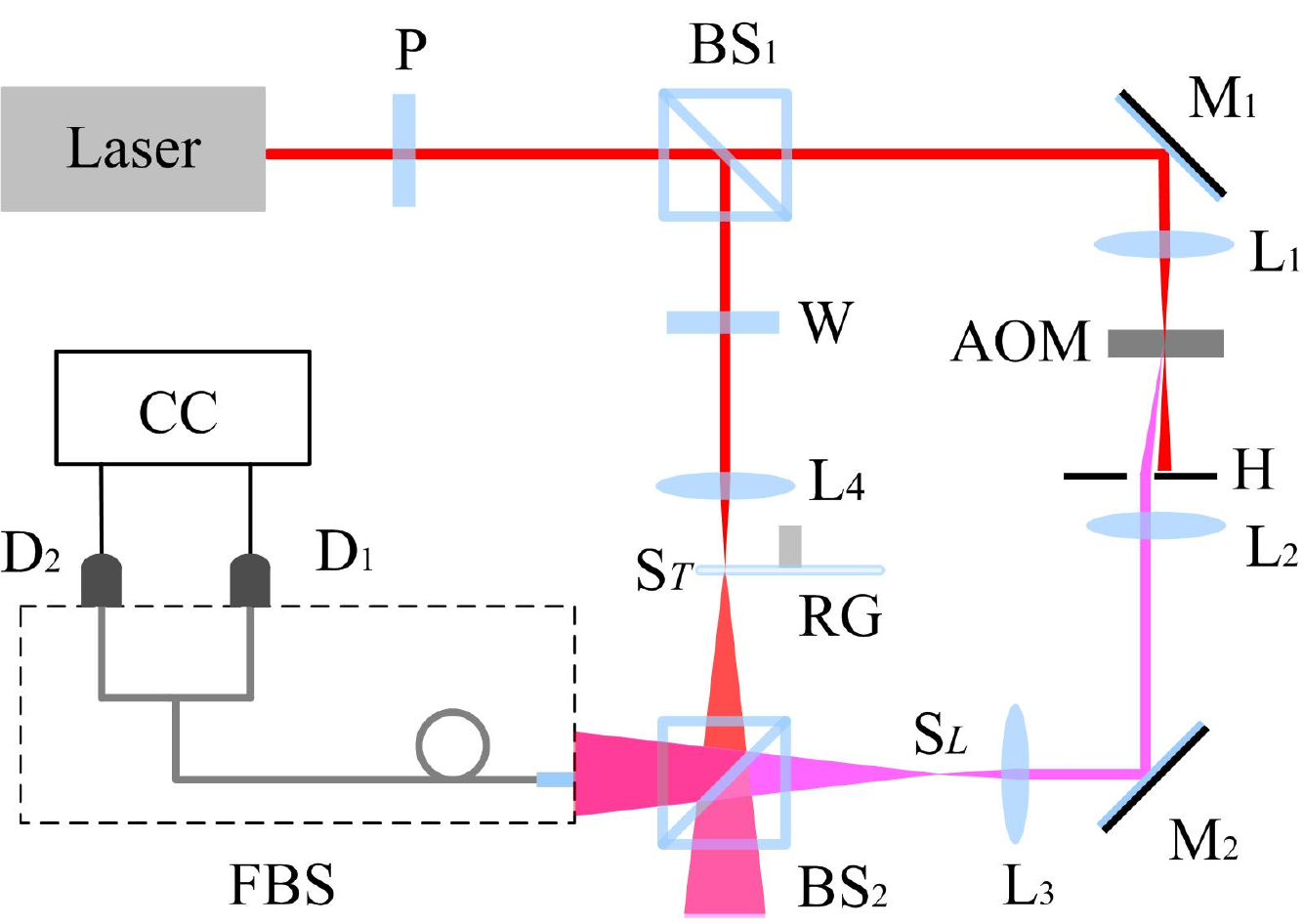}
    \caption{The experimental setup for the first- and second-order interference of pseudothermal and laser light. Laser: 780 nm single-mode laser with bandwidth of 200 kHz. P: Polarizer. BS: 1:1 nonpolarized beam splitter. W: $\lambda/2$ wave plate. RG: Rotating ground glass. S: Light source. L: Lens. M: Mirror. AOM: Acoustooptic modulator. H: Pinhole. FBS: Fiber beam splitter. D: Single-photon detector. CC: two-photon coincidence count detection system. See text for details.}\label{experiment}
\end{figure}

In the last section, we have calculated the first- and second-order temporal interference between thermal and laser light shown in Fig. \ref{setup}(a) and (b), respectively. It is concluded that the first-order interference pattern can not be observed, while the second-order interference pattern can be observed. In this section, we will employ experimental scheme in Fig. \ref{experiment}, which is similar as our earlier experimental setup in Ref. \cite{liu-arXiv}, to verify the predictions. The laser is a single-mode continuous wave laser with 780 nm central wavelength and 200 kHz frequency bandwidth. P is a polarizer. BS$_1$ and BS$_2$ are 1:1 nonpolarized beam splitters. W is a $\lambda/2$ wave plate to control the polarization. RG is Rotating ground glass to randomize the phases of photons passing through it. M$_1$ and M$_2$ are mirrors. L$_1$ and L$_2$ are two identical lens with focus length of 100 mm and the distance between them are 200 mm. Acoustooptic modulator (AOM) is at the confocal point of L$_1$ and L$_2$ to change the frequency of laser light. H is a pinhole to block the laser light that does not change frequency after passing through AOM. L$_3$ and L$_4$ are two identical lens with focus length of 50 mm. S$_T$ and S$_L$ are point pseudothermal and laser light sources, respectively. FBS is a 1:1 nonpolarized fiber beam splitter. The distance between L$_3$ and the collector of FBS is equal to the distance between L$_4$ and the collector of FBS via BS$_2$. The optical length between the laser and detector via M$_1$ is 4.24 m. The single-photon counting rates of D$_1$ and D$_2$ are both about 50000 c/s, which means on average there is only $1.41 \times 10^{-3}$ photon in the experimental setup at one time. Our experiments are done at single photon's level.

\begin{figure}[htb]
    \centering
    \includegraphics[width=70mm]{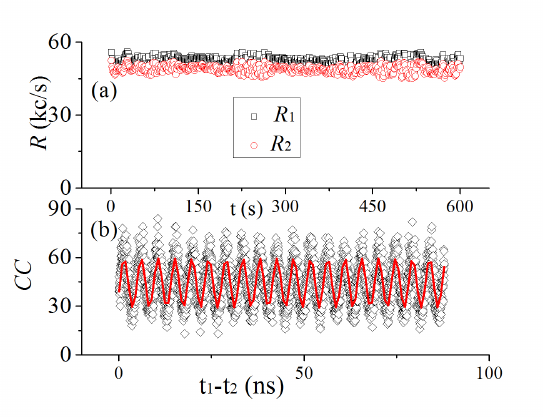}
    \caption{The observed first- and second-order temporal interference patterns. The first- and second-order temporal interference patterns are shown in (a) and (b), respectively. $R_1$ and $R_2$ in (a) are single-photon counting rates of $D_1$ and $D_2$, respectively. $CC$ is two-photon coincidence counts for 600 s. $t_1-t_2$ is the time difference between the two single-photon detection events within a two-photon coincidence count. The data in (a) and (b) is recorded simultaneously. }\label{beating1}
\end{figure}

\begin{figure}[htb]
    \centering
    \includegraphics[width=70mm]{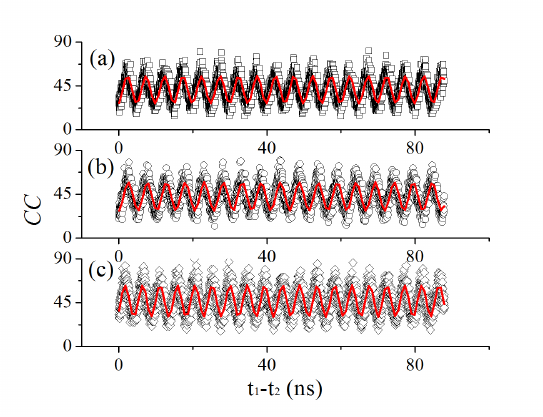}
    \caption{The second-order temporal beatings when the frequency shifts of AOM are 212.51 MHz, 200.66 MHz, and 195.28 MHz for (a), (b) and (c), respectively.}\label{beating2}
\end{figure}

We first measure the first- and second-order temporal interference patterns when the $\lambda/2$ wave plate, W, is removed. The observed first- and second-order interference patterns are shown in Fig. \ref{beating1}(a) and (b), respectively. The dark counts of both detectors are less than 100 c/s. The single-photon counting rates of D$_1$  and D$_2$ are shown by the squares and circles in Fig. \ref{beating1}(a), respectively. No first-order temporal interference pattern is observed by either D$_1$ or D$_2$, which is consistent with the prediction of Eq. (\ref{1st-3}). In the same condition, the second-order temporal interference pattern is observed in Fig. \ref{beating1}(b), which is consistent with the prediction of Eq. (\ref{G2-4}). The reason why the background of the observed second-order temporal beating is flat is the second-order coherence time of pseudothermal light is much longer than the beating period. The second-order coherence time of pseudothermal light is measured to be 51 $\mu$s in our experiment. The beating period in Fig. \ref{beating1}(b) is 4.85 ns. Figure \ref{beating2}(a), (b), and (c) correspond to the second-order temporal beatings when the frequency shifts of AOM are 212.51 MHz, 200.66 MHz, and 195.28 MHz, respectively. The calculated beating frequencies are 212.04 MHz, 199.92 MHz, 194.86 MHz, respectively, which are consistent with the frequency shifts of AOM.

\begin{figure}[htb]
    \centering
    \includegraphics[width=70mm]{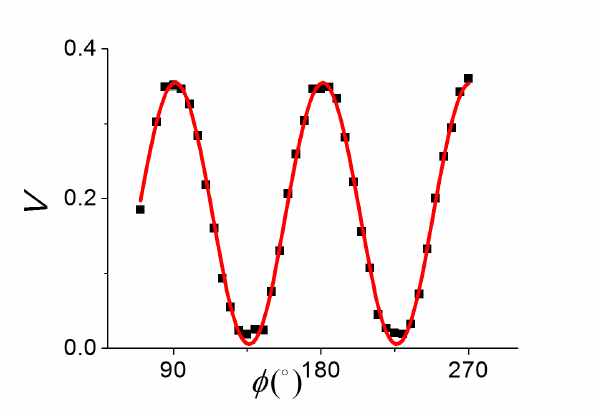}
    \caption{Visibility of the second-order temporal beating versus the angle of $\lambda/2$ wave plate.}\label{visibility}
\end{figure}

We also measured the second-order temporal beating when the polarizations of thermal and laser light are different. The visibility of the second-order temporal beating is shown in Fig. \ref{visibility} when the angle of W is varied. The observed maximum visibility is $35.12(\pm 0.63)$\% when the polarizations of these two light beams are parallel. The observed minimum visibility is $1.88(\pm 0.46)$\% when the polarizations of these two light beams are orthogonal, which approaches 0. The reasons why the visibility can not reach zero may be the polarization of the light beam is not 100\% polarized in one direction and the polarizations of these two light beams are not strictly orthogonal in the measurement.

\section{Discussions}\label{discussions}

Although the first- and second-order interference of classical light can be interpreted by both quantum and classical theories \cite{glauber-1,glauber-2}, we employed one-photon and two-photon interference based on the superposition principle in Feynman's path integral theory. Not only because it is simple, but also it will give a unified interpretation for all order interference of classical and nonclassical light. In classical theory, the first-order interference of light is interpreted by the superposition principle of electromagnetic fields \cite{born}. The second- and higher-order interference of light is interpreted by the intensity fluctuation correlations based on the first-order interference of light \cite{hbt-1957,hbt-1958}. The first-order interference of light is the foundation of the second- and higher-order interference of light in classical theory. In quantum theory, the first-order interference of light is interpreted by one-photon interference based on the superposition principle in quantum physics. The second- and higher-order interference is interpreted by multi-photon interference based on the same superposition principle in quantum physics. The first-, second-, and higher-order interference of light is interpreted by the same theory in a unified way \cite{feynman}. Further more, the superposition principle in Feynman's path integral theory can be easily generalized to interpret all order interference of massive particles, such as electrons, neutrons, and atoms. The classical theory of interference of light, on the other hand, can not be generalized to the interference of massive particles or nonclassical light.

\begin{figure}[htb]
    \centering
    \includegraphics[width=70mm]{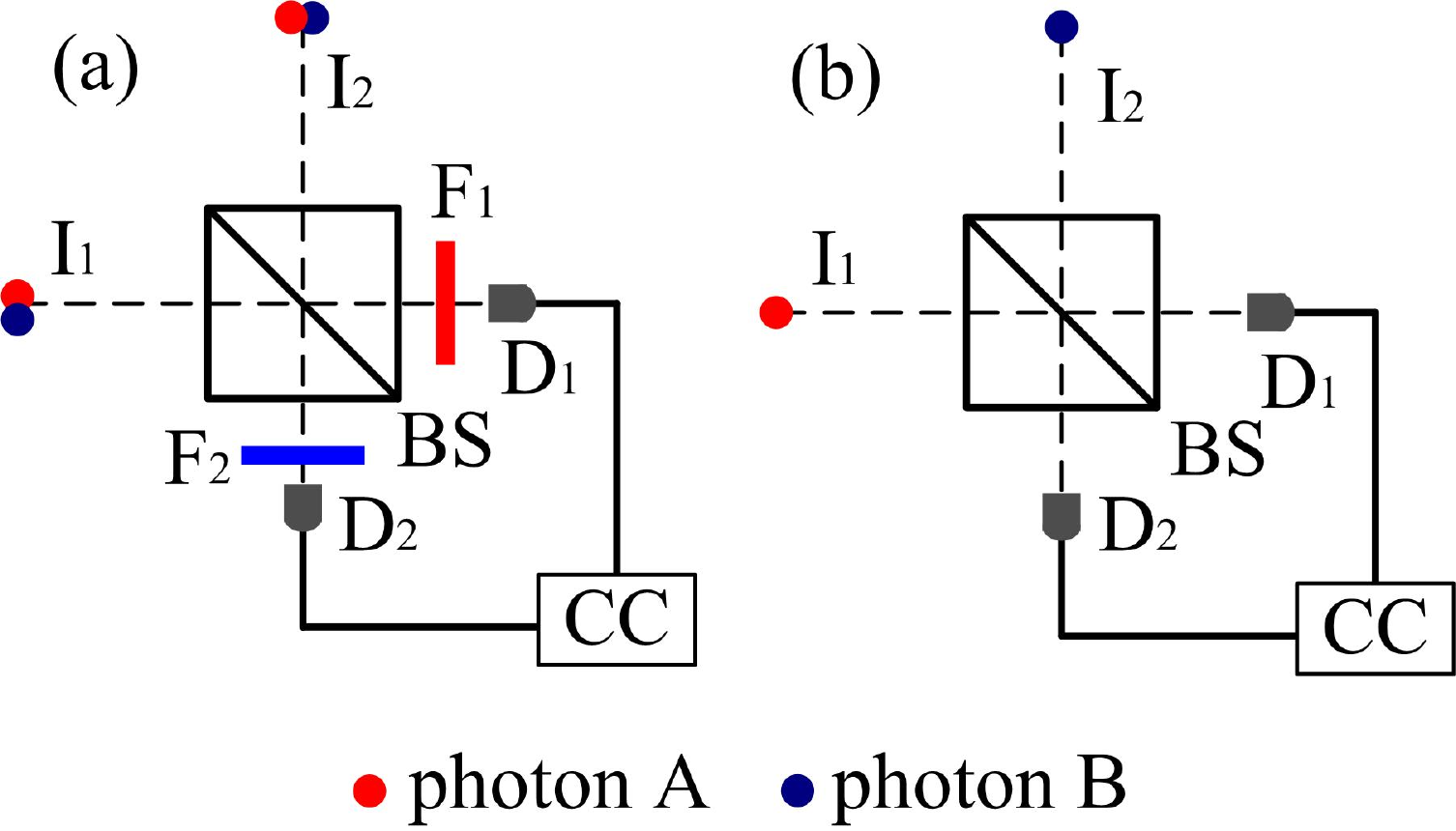}
    \caption{Two different situations for two photons in a HOM interferometer.I: input port. F: Filter. Other symbols are the same as the ones in Figs. \ref{setup} and \ref{experiment}.}\label{discussion}
\end{figure}

The superposition principle in Feynman's path integral theory is based on the indistinguishability of different alternatives \cite{feynman}. The indistinguishability of alternatives is related, but not equivalent, to the indistinguishability of particles. For instance, there are two different situations for two photons in a Hong-Ou-Mandel (HOM) interferometer as shown in Fig. \ref{discussion}. I$_1$ and I$_2$ are two input ports, respectively. F$_j$ is frequency filter that only let photon with frequency $\omega_j$ passes ($j=1$ and 2).  In Fig. \ref{discussion}(a), D$_1$ and D$_2$ can only be triggered by photons A and B, respectively. For simplicity, we only consider the case that one photon comes from one input port, respectively. There are two different alternatives to trigger a two-photon coincidence count in Fig. \ref{discussion}(a). The first one is photon A coming from I$_1$ is detected by D$_1$ and photon B coming from I$_2$ is detected by D$_2$. The second one is photon A coming from I$_2$ is detected by D$_1$ and photon B coming from I$_1$ is detected by D$_2$. Although photons A and B are distinguishable, these two different alternatives are indistinguishable if it is impossible to tell which photon comes from which input port.  It is the reason why the beating between photons of different colors can be observed with ordinary detectors \cite{ou-1988,ou-1988-oc}. In the scheme shown in Fig. \ref{discussion}(a), the indistinguishability of alternatives is not equivalent to the indistinguishability of photons.

In the scheme shown in Fig. \ref{discussion}(b), photons A and B come from I$_1$ and I$_2$, respectively. There are no filters before detectors. There are two different ways to trigger a two-photon coincidence count, which are $A\rightarrow D_1, B\rightarrow D_2$ and $A\rightarrow D_2, B\rightarrow D_1$, respectively. If these two photons are distinguishable for the detection system, these two alternatives are distinguishable. If these two photons are indistinguishable, these two different ways are indistinguishable, too. The indistinguishability of alternatives is equivalent to the indistinguishability of photons in the scheme shown in Fig. \ref{discussion}(b).

Our experimental setup is similar as the one in Fig. \ref{discussion}(b) except there are more than two alternatives to trigger a two-photon coincidence count. The second-order temporal beating between photons of different frequencies is observed as the ones in Figs. \ref{beating1} and \ref{beating2}. The reason why there are two-photon interference for photons of different frequencies is photons with different frequencies can be indistinguishable \cite{liu-arXiv} (and references therein). These two different alternatives to trigger a two-photon coincidence count are indistinguishable and there is two-photon interference \cite{feynman}. When the polarization of pseudothermal light is changed by rotating the $\lambda/2$ wave plate, the photons in these two light beams gradually become distinguishable. The visibility of the second-order temporal beating drops from $35.12(\pm 0.63)$\% to nearly zero when the polarizations of pseudothermal and laser light change from parallel to orthogonal. When the polarizations of the photons in these two light beams are orthogonal, although these photons are indistinguishable by frequencies, they are distinguishable by polarizations. These two different alternatives to trigger a two-photon coincidence count are distinguishable. There is no two-photon interference and no second-order interference pattern can be observed in this condition.

\section{Conclusions}\label{conclusions}

In conclusions, we have discussed the first- and second-order temporal interference between thermal and laser light based on the superposition principle in Feynman's path integral theory. It is concluded that the first-order interference pattern can not be observed by superposing thermal and laser light, while the second-order temporal beating can be observed in the same condition. These predictions are experimentally verified by employing pseudothermal light to simulate thermal light. The relationship between the indistinguishability of alternatives and the indistinguishability of photons is dependent on the employed experimental schemes. The conclusions in this paper can be generalized to the third- and higher-order interference of light. By changing the Feynman's propagators and superposition principle for fermions, the same method can be employed to calculate the first-, second- and higher-order interference of massive particles.

\section*{Acknowledgments}
The authors wish to thank D. Wei for the help on the AOM. This project is supported by National Science Foundation of China (No.11404255), Doctoral Fund of Ministry of Education of China (No.20130201120013), the 111 Project of China (No.B14040) and the Fundamental Research Funds for the Central Universities.

\end{document}